# Anisotropic Magnetoresistance in $Ga_{1-x}Mn_xAs$ films

*K.Y. Wang, K.W. Edmonds, R.P. Campion, L.X. Zhao, A.C. Neumann, C.T. Foxon, B.L. Gallagher, P.C. Main, and C.H. Marrows*[1]
*School of Physics and Astronomy, University of Nottingham, Nottingham NG7 2RD, UK*
[1]*Department of Physics, University of Leeds, Leeds LS2 9JT*

Abstract

The magnetoresistance in a series of $Ga_{1-x}Mn_xAs$ samples with $0.02 \le x \le 0.08$ have been measured for three mutually orthogonal orientations of the external magnetic field. We find that in high quality GaMnAs epilayers, both $AMR_{//}$ and $AMR_{\perp}$ effects decrease with increasing Mn concentration. This is most probably due to the increase in the density of compensating defects evident from the measured hole densities. The sign of the AMR and the dependence of the ratio of $AMR_{\perp}/AMR_{//}$ on strain are in agreement with theoretical predictions. We show that the dependence of the measured AMR on the applied field can be used to obtain information on the dependence of the magnetisation on the applied field. We find that the sample studied have strong in plane anisotropy, as is expected for compressively strained GaMnAs films, but also a clear uniaxial anisotropy with [110] the easy axis.

Introduction:

The recent discovery of ferromagnetic semiconductor $Ga_{1-x}Mn_xAs$ films with the Curie temperature as high as 110 K [1] has attracted considerable attention due to their potential for use in novel devices making use of both magnetic and conventional semiconductor properties. The transport and optical properties of these heavily-doped semiconductors are richer than those of conventional itinerant electron ferromagnets because of strong valence-band spin-orbit coupling.

In this system Mn atoms incorporated into a GaAs matrix play two roles: they act as both s=5/2 local moments, and as acceptors generating hole states in the material [2]. When GaMnAs epilayers are grown on GaAs substrates the lattice mismatch results in the GaMnAs layer being compressively strained. High resolution X-ray diffraction studies have shown that films of (Ga,Mn)As several microns thick, grown on GaAs, are still fully compressively strained [3]. It has been demonstrated that this compressive strain results in a strong in-plane magneto-crystalline anisotropy that is much larger than the shape anisotropy [4,5]. Recently it has been predicted that such GaMnAs epilayers should also show large anisotropic magnetoresistance (AMR) and that the magnitude and sign of the AMR will be dependent on the strain [6].

In this paper, we report the observation of large AMR in high quality MBE grown GaMnAs layers. We show that the sign and strain dependence of the AMR is consistent with the predictions of Boltzmann transport calculations [6]. We also demonstrate that the measured AMR can be used to provide information about the magnetic anisotropy. We find that the GaMnAs layers have a clear in-plane uniaxial easy axis along [110].

Experimental and discussion:

The $Ga_{1-x}Mn_xAs$ ($0.02 \leq x \leq 0.08$) films were grown on semi-insulating (001) GaAs substrates by low temperature MBE. Prior to $Ga_{1-x}Mn_xAs$ deposition, a 100-nm GaAs buffer layer was grown at $580^0$. Then the substrate was cooled down to $175^0$-$240^0$C for the growth of a 50 nm thick low-temperature GaAs layer, followed by a 50 nm thick $Ga_{1-x}Mn_xAs$ layer. The lattice constant, a, of the $Ga_{1-x}Mn_xAs$ layers was found to follow Vegard's law, with a = 5.6545*(1-x)+5.90*x. The $Ga_{1-x}Mn_xAs$ films are fully compressive strained. The strained films are characterized by a substrate-film lattice mismatch [7] $e=(a_{sub}-a_{film})/a_{film}$, where $a_{sub}$ and $a_{film}$ are the respective relaxed lattice constant of substrate and film, and the results are shown in Table 1.

The longitudinal resistance $R_{xx}$ and Hall resistance $R_{xy}$ were measured using a standard low-frequency lock-in technique. The anisotropic magnetoresistance results presented in this paper are for normal Hall bars and L shaped samples in which it is possible to measure the AMR for the current in either the [110] or $[1\bar{1}0]$ direction. In discussing the results for the L shaped samples the current direction is defined to be x direction, the y direction is in the plane and perpendicular to the current direction, and the film growth direction [001] direction is z direction.

The Hall resistivity in magnetic materials is empirically known to be a sum of an ordinary Hall term and an anomalous Hall term as [8]:

$$R_{Hall} = \frac{R_0}{d}B + \frac{R_S}{d}M \qquad (6)$$

where $R_0$ is the ordinary Hall coefficient, $R_S$ the anomalous Hall coefficient, d the sample thickness, and M the magnetization of the samples. $R_S$ is proportional to $R_{xx}^{\gamma}$ with a constant of proportionality that is usually assumed to be temperature and magnetic field independent. A detailed study of the Hall effect in our samples [9,10] at high magnetic fields indicates that $\gamma$ generally lies between 1 and 2, and also yields the values for the hole densities given in Table 1.

Because of the high-hole concentration, the contribution from the ordinary Hall term is negligible at low fields for most of the investigated temperature range. The ferromagnetic transition temperature $T_C$ can be determined from the field dependence of magnetization $M_{Hall}$ obtained from magnetotransport measurement ($M_{Hall} = dR_{xy}/dR_{xx}$), using an Arrott plot ($M^2$ vs B/M). The results are shown in Table 1. The highest $T_C$, which resulted from a post growth annealing, is 112 K which is the same as the highest $T_C$ achieved so far.

As can be seen from figure 1, the observed AMR is quite complex. For large applied fields the AMR saturates and an isotropic negative magnetoresistance is observed. This "high field AMR" occurs for external fields at which the magnetisation is saturated in the direction of the applied magnetic field. Below the saturation fields the behaviour is determined by the form of $M(B_{ext})$. This "low field AMR" is discussed later.

Fig. 1 shows that at high fields the resistance is always lower for the applied field perpendicular to the current than for the applied field parallel to the current. The behaviour is opposite to that normally observed in ferromagnets [11] but agrees with other observations for this system [12,13] and with the predictions of the mean field theory [6].

Fig.2 shows the measured $AMR_{//} = (R_{//x}-R_{//y})*100/R_{//x}(\%)$ and $AMR_{\perp} = (R_{//x}-R_{//z})*100/R_{//x}(\%)$. Both the $AMR_{//}$ and the $AMR_{\perp}$ decreases with increasing Mn concentration. The high field $AMR_{//}$ and $AMR_{\perp}$ are, within error, equal for the x = 2% sample. The high field ratio of $AMR_{\perp}/AMR_{//}$ increases with increasing Mn content. Theoretical calculations [6] predict that the ratio $AMR_{\perp}/AMR_{//}$ should increase as the magneto-crystalline anisotropy field increases, with increasing compressive stain. These results are consistent with this prediction since there is very little strain for the sample with x = 2% but increasing compressive strain at higher Mn concentrations. The decrease of the magnitude of the AMR with increasing of Mn content is most probable due to the increase in the concentrations of compensating defects [9].

The $AMR_{//}$ drops very quickly at low external magnetic field and then saturates. The form of $AMR_{\perp}$ is more complicated as it results from the very different magnetic field dependences of $R_{//z}$ and $R_{//x}$, which are discussed in detail below.

When the current direction is along [110] a large increase in resistance is observed as the magnitude of the applied magnetic field increases, for $\theta = 90^0$ (see Fig. 1a). For $\theta = 0^0$ the corresponding change in resistance is much smaller while the curves for $\theta = 45^0$ and $-45^0$ almost coincide. When the current direction is along $[1\bar{1}0]$ a large decrease in resistance is observed as the magnitude of the applied magnetic field increases, for $\theta = 0^0$ (see Fig. 1c) while for $\theta = 90^0$ the corresponding change in resistance is much smaller.

If the domains in the samples were randomly orientated for zero external field then the anisotropic part of the magnetoresistance would be zero for $\theta = 45^0$ and $-45^0$ and the magnitude of the resistance change for $\theta = 0^0$ and $\theta = 90^0$ would be equal. Fig 1 therefore shows that the magnetisation of the magnetic domains are preferentially oriented along [110] direction in the absence of an applied magnetic field. This means that most. The observed change in resistance is small when the applied external field is along the [110] direction as the magnetisation already lies along [110]. When the applied external field is along the $[1\bar{1}0]$ direction the observed resistance change is largest as the magnetisation rotates out of the easy [110] direction into the [110] direction.

Fig.1b shows that the resistance increases much more slowly with B when the external magnetic field is perpendicular to the film (B//z) than when it is the plane of the film[1].

[1] Fig 1d shows similar behaviour but in this case the AMR change for B//z is small (comparable with the isotropic magnetoresistance), as the magnetisation lie along y (i.e. perpendicular to the current) without a magnetic field.

This is because the anisotropy in the plane of the film is much stronger than the anisotropy between different directions within the plane of the film.

In order to show more clearly the functional form of R(B) for the different external field directions we first subtracted the isotropic part of the magnetoresistance from the data of fig. 1b. The normalised AMR, ((R(B)-R(B=0))/((R(B=0.7T)-R(B=0)) is then plotted in figure 4. This figure shows that the normalised AMR reaches a maximum for different magnetic field values. This field, which give a measure of the strength of the anisotropy, is smallest for B//y ([110]}, which we have already established is the easy uniaxial direction. It is a little larger for B//x and much larger for B//z.

The functional form of $AMR_{\perp}$ shown in figure 2 is due to the difference between $R_{//y}(B)$ and $R_{//x}(B)$ which both have similar functional forms. The functional form of $AMR_{\perp}$ is due to the difference between $R_{//z}(B)$ and $R_{//x}(B)$ which both have very different functional forms. The sharp drop of $AMR_{\perp}$ at low field is due to the $R_{//x}$ decreases at low magnetic field while the shoulder at larger B comes from the more slowly varying $R_{//z}$.

We conclude that the samples have strong in plane anisotropy, as is expected for compressively strained GaMnAs films, and weaker uniaxial anisotropy within the plane.

As figure 3 shows the uniaxial anisotropy is also observed in vibrating sample magnetometer measurements on these samples.

Conclusion:

We find that in high quality GaMnAs epilayers, both $AMR_{//}$ and $AMR_{\perp}$ effects decrease with increasing Mn concentration. This is most probably due to the increase in the density of compensating defects evident from the measured hole densities. The sign of the AMR and the dependence of the ratio of $AMR_{\perp}/AMR_{//}$ on strain are in agreement with theoretical predictions [6]. We have shown that the dependence of the measured AMR on the applied field can be used to obtain information on the dependence of the magnetisation on the applied field. We find that the sample studied have strong in plane anisotropy, as is expected for compressively strained GaMnAs films, but also a clear uniaxial anisotropy with [110] the easy axis.

Acknowledgements:

The project was funded by EPSRC and the European Union. Thanks to Jaz Chauhan and Dave Taylor for processing of Hall bars.

Table I: Measured properties of the $Ga_{1-x}Mn_xAs$ films: Mn concentration, compressive strain, $\sigma$ , hole density, p, Curie temperature, $T_c$, Mobility at 4.2K, $\mu$, and anisotropic magnetoresistance for B = 0.75T. The uncertainty in the compresive strain, hole density, mobility and Curie Temperature is about 5%. The Mn concentration is the nominal one aimed for in growth which agrees within ~10% with that obtained from X-ray and electron microprobe determinations

| No | Mn (%) | $\sigma$ (%) | p ($m^{-3}$) | $T_C$ (K) | $\mu$ ($m^2V^{-1}s^{-1}$) | $(R_{//y}-R_{//x})/R_{//x}$ | $(R_{//z}-R_{//x})/R_{//x}$ |
|---|---|---|---|---|---|---|---|
| 22 | 2 | 0.087 | 3.96 x$10^{26}$ | 60 | 3.96 x$10^{-4}$ | 6.4 ± 0.2 % | 6.4 ± 0.2 % |
| 21 | 3 | 0.13 | 4.24 x$10^{26}$ | 63 | 4.20 x$10^{-4}$ | 5.4 ± 0.2 % | 5.9 ± 0.2 % |
| 18 | 5 | 0.22 | 5.01 x$10^{26}$ | 76 | 3.77 x$10^{-4}$ | 3.4 ± 0.2 % | 4.2 ± 0.2 % |
| 24 | 6 | 0.26 | 4.84 x$10^{26}$ | 76 | 3.32 x$10^{-4}$ | 2.3 ± 0.2 % | 3.8 ± 0.2 % |
| 31 | 8 | 0.35 | 5.01 x$10^{26}$ | 111 | 2.27 x$10^{-4}$ | 0.9 ± 0.2 % | 1.6 ± 0.2 % |

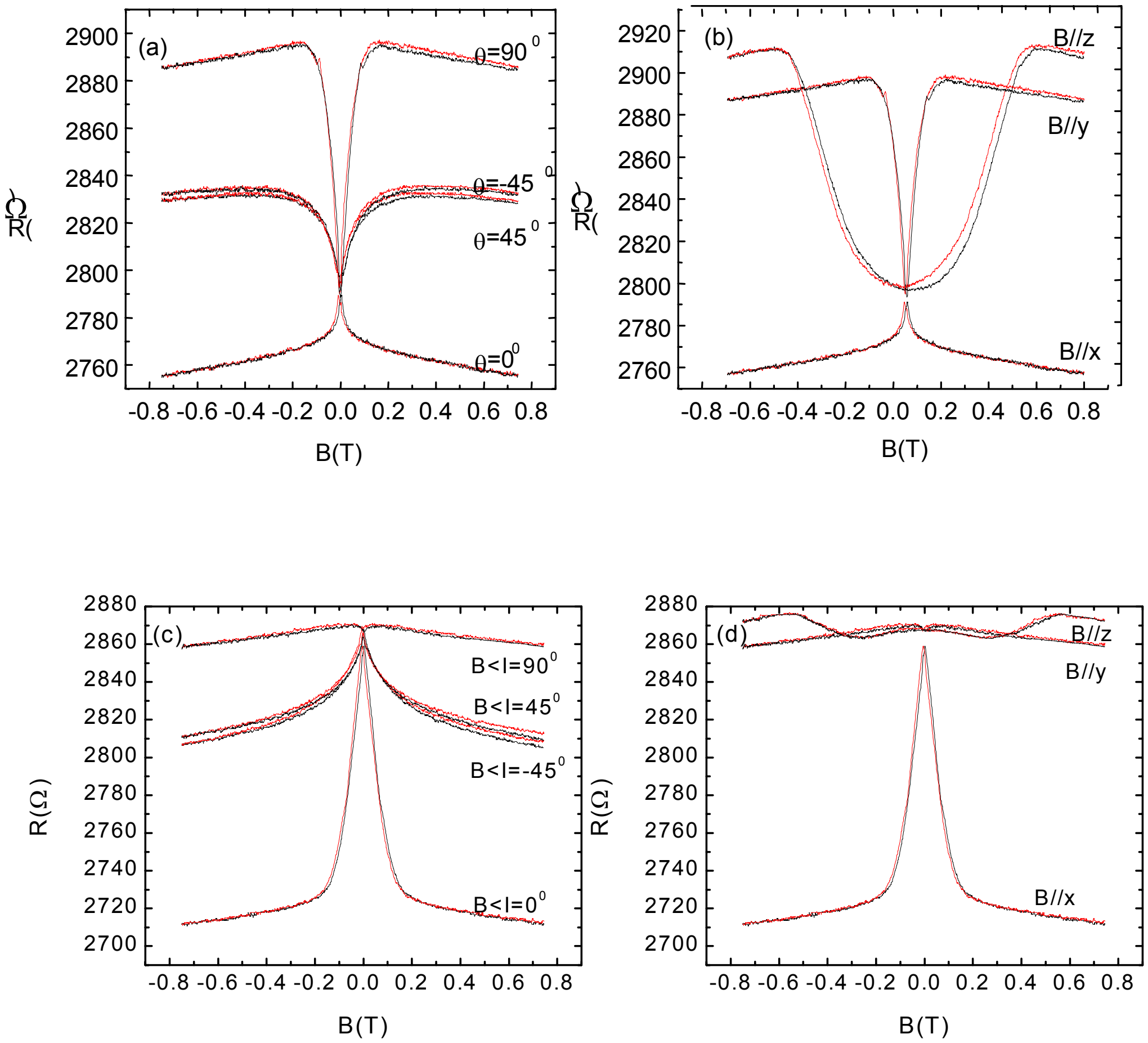

Fig 1. Measured magnetoresistance for an L shaped sample of GaMnAs with 3% Mn measured at 4.2K. (a) Current along [110] with the magnetic field in plane of the film at an angles φ to the current direction. (b) Current along [110], See text for definition of axes. (c) Current along [$1\bar{1}0$] with the magnetic field in plane of the film at an angles φ to the current direction (d) Current along [$1\bar{1}0$] see text for definition of axes.

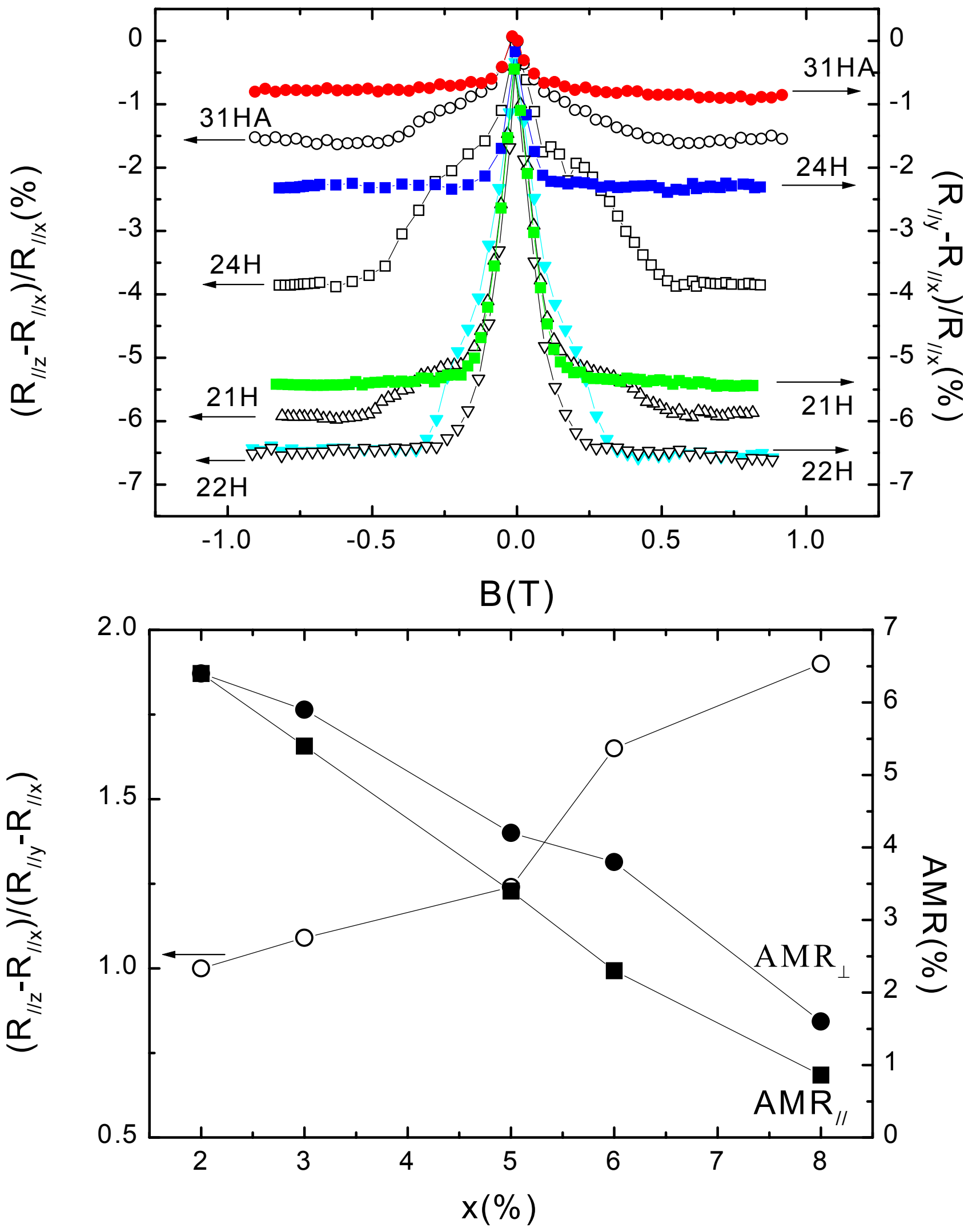

Fig. 2: (a)The AMR$_\perp$, AMR$_{//}$ vs. B for the $Ga_{1-x}Mn_xAs$ samples with 2, 3, 6 and 8%Mn (b) The high field AMR$_\perp$, AMR$_{//}$ and the ratio of AMR$_\perp$/AMR$_{//}$ for the $Ga_{1-x}Mn_xAs$ samples with different Mn concentration.

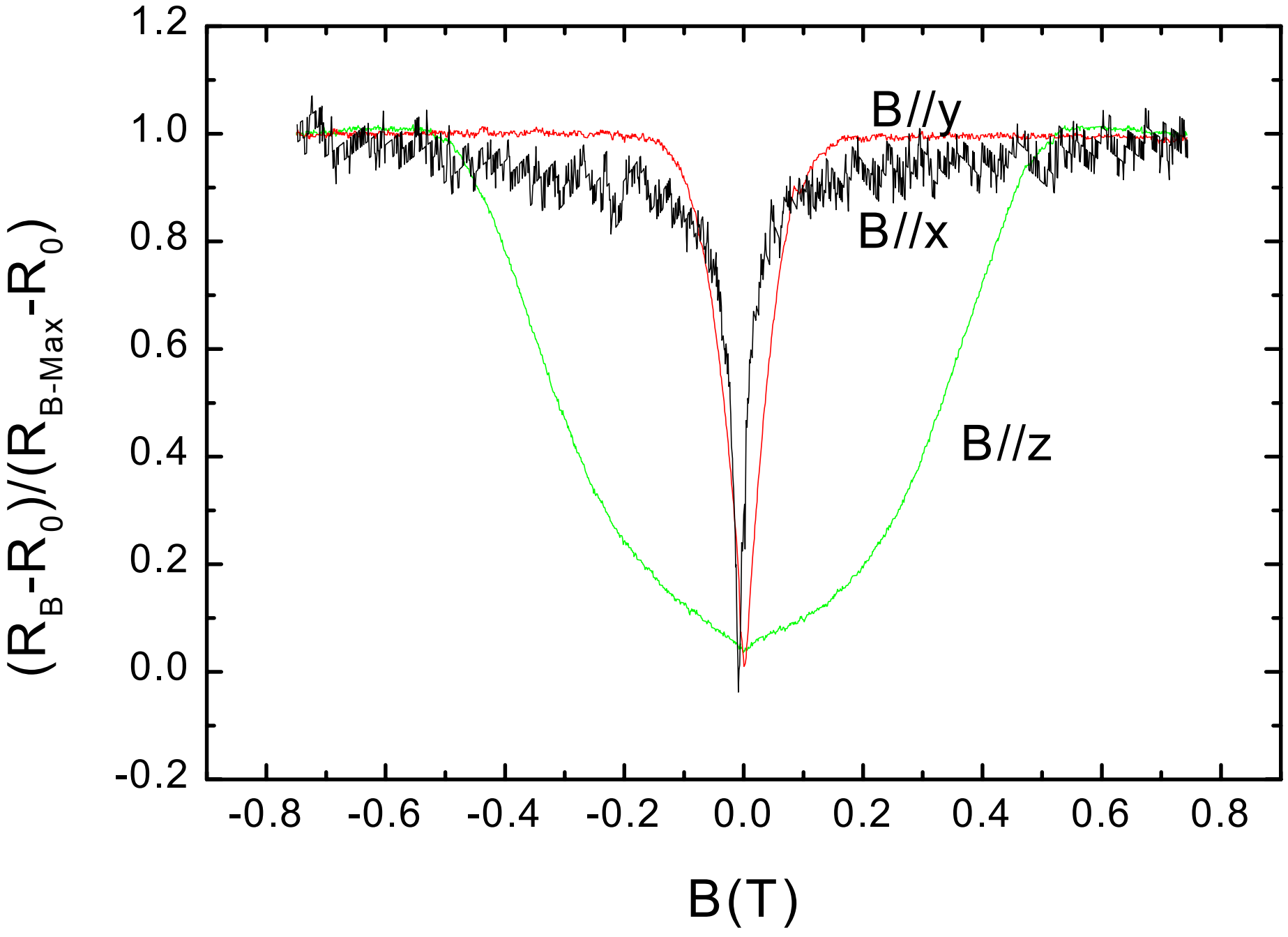

Fig 3. The normalised AMR obtained from the data of Fig 1b after subtraction of the isotropic magnetoresistance.

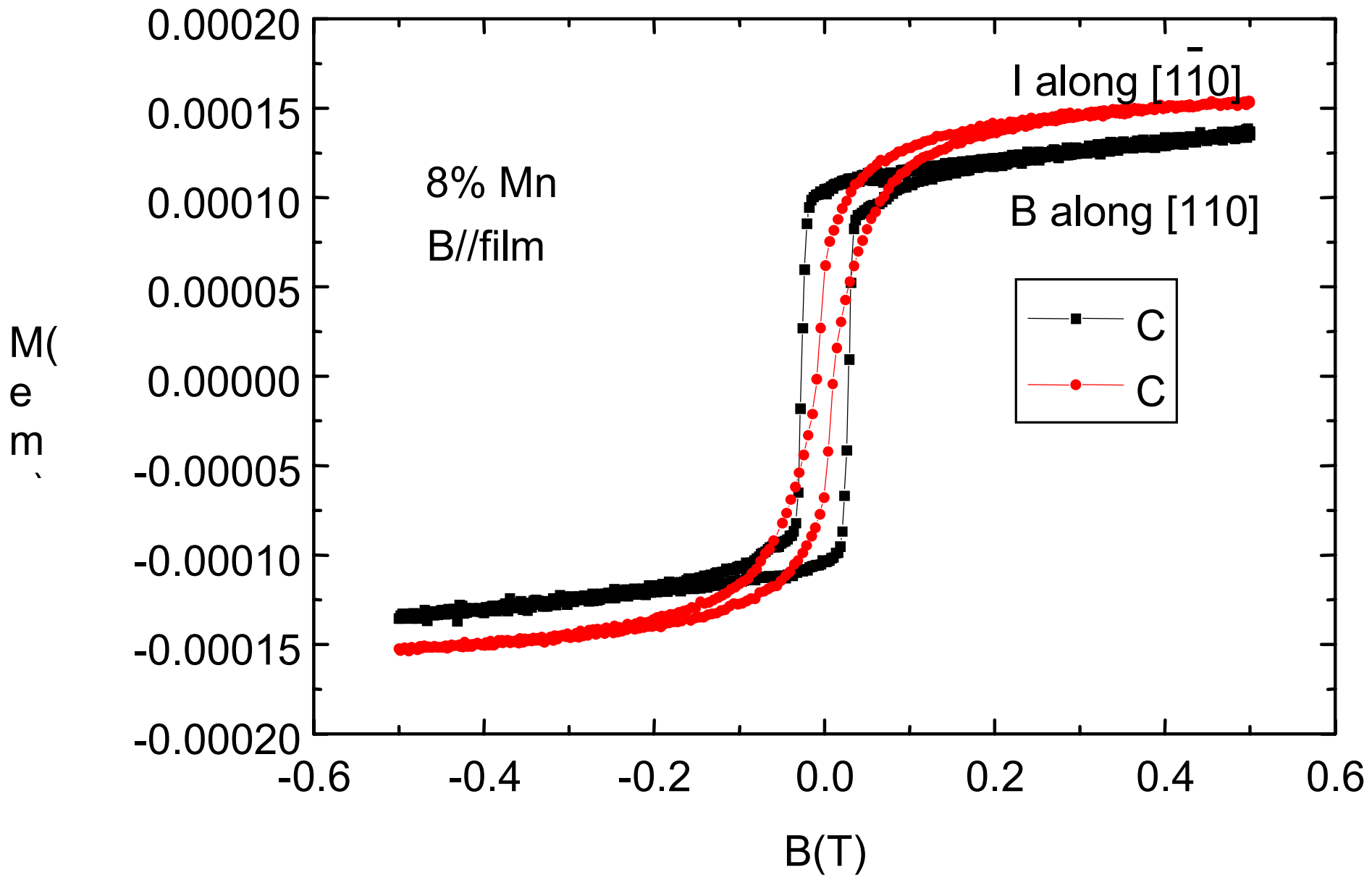

Fig. 4. VSM results for a 8% Mn sample.